\documentclass[prd,superscriptaddress,showpacs,showkeys,preprint]{revtex4}
\usepackage{epsfig}

\newcommand{\fms}[1]{{#1}\!\!\!/}
\newcommand{\fmsl}[1]{{#1}\!\!\!\!/}

\newcommand{\n}{\overline{n}}
\newcommand{\oL}{\bar{\Lambda}}

\begin{document}
\title{SCET sum rules for heavy-to-light form factors}
\author{Junegone Chay}\email{chay@korea.ac.kr}
\affiliation{Department of Physics, Korea University, Seoul 136-701, Korea} 
\author{Chul Kim}\email{chk30@pitt.edu}
\affiliation{Department of
  Physics and Astronomy, University of Pittsburgh, PA 15260, U.S.A.}
\author{Adam K. Leibovich}\email{akl2@pitt.edu}
\affiliation{Department of
  Physics and Astronomy, University of Pittsburgh, PA 15260, U.S.A.}

\date{\today\\ \vspace{1cm} }
\preprint{}
\baselineskip 3.0ex 
\parskip 2.0ex

\begin{abstract}
\baselineskip 3.0ex 
We consider a sum rule for heavy-to-light form factors in
soft-collinear effective theory (SCET). Using the correlation function
given by the time-ordered product of a heavy-to-light current and its
hermitian conjugate, the heavy-to-light soft form factor $\zeta_{P}$
can be related to the leading-order $B$ meson shape function. 
Using the scaling behavior of the heavy-to-light form factor in
$\Lambda_{\rm   QCD}/m_b$, we put a constraint on the behavior of the
$B$ meson shape function near the endpoint. We employ the sum rule
to estimate the size of $\zeta_{P}$ with the model for the shape
function and find that it ranges from 0.01 to 0.07.
\end{abstract}

\maketitle

Information on heavy-to-light form factors is important to
extract Standard Model parameters from
experimental results.  For instance, 
the $B\to\pi$ form factor is needed to extract $V_{ub}$ or
the CKM angle $\gamma$ from exclusive $B$ decays
\cite{gammaextraction}.  Due to the non-perturbative nature of the form
factor, techniques beyond perturbation theory need to be employed to
try to evaluate these functions.  Unquenched lattice results are starting to
be available for the $B\to\pi$ form factor \cite{lattice}.  Due to the
uncertainties in the lattice technique, the pion is restricted to have
energies $E_\pi \lesssim 1$ GeV, just in the region where the
experimental uncertainty is largest.  Another method for determining
the form factor is light-cone sum rules \cite{lcsr}.  In
Ref.~\cite{DeFazio:2005dx} (see also \cite{Ball:2003bf}), light-cone
sum rules were investigated using soft-collinear effective theory
(SCET) \cite{scet}, where it is argued that using SCET
allows for a more consistent scheme to calculate both factorizable and
non-factorizable contributions than the traditional light-cone sum
rule approach. 
 
In this paper, we investigate a novel sum rule using SCET, relating
the $B\to\pi$ form factor at small $q^2$ to the $B$ meson
shape function \cite{shape} which describes the motion of the $b$
quark inside the $B$ meson.  The shape function cannot be calculated,
but it can be 
extracted from the data \cite{Bornheim:2002du}.  After relating the
two non-perturbative functions using the sum rule, we use the scaling
in $\Lambda_{\rm QCD}/m_b$ of the $B\to\pi$ form factor to put
constraints on shape function models near the endpoint
region. By choosing a model for a shape function which satisfies the
constraints, we can give model-dependent values for the $B\to\pi$ form
factor at $q^2=0$.

\begin{figure}[t]
\begin{center}
\epsfig{file=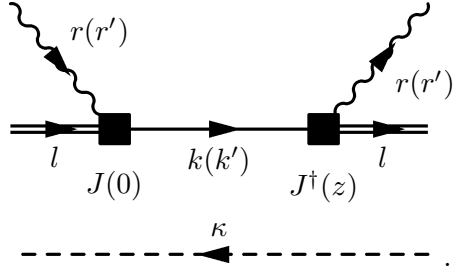, width=6cm}
\end{center}
\vspace{-1.0cm}
\caption{Feynman diagram for the correlation function  $\Pi$
in the timelike (spacelike) region. 
The double line is a heavy quark field, the solid line is a
hard-collinear quark with the residual
momentum $k (k')$. The dashed line is a spectator quark in the $B$
meson, which is an usoft field in $\rm{SCET_I}$.} 
\label{fig1}
\end{figure}

Let us consider the correlation function in $\rm{SCET_I}$
\cite{Bauer:2002aj}, defined as 
\begin{equation} 
\Pi(q) = i \int d^4 z e^{iq\cdot z} \langle \overline{B} |\mathrm{T} 
[J_0^{\dagger} (z) J_0 (0)] | \overline{B} \rangle, 
\label{cor} 
\end{equation} 
where the leading-order heavy-to-light current in SCET can be written as 
\begin{equation} 
J_0(x) = e^{i(\tilde p - m_b v)\cdot x} \overline{\xi}_n W
Y^{\dagger} h_v (x).
\label{cur} 
\end{equation} 
Here $\tilde p = \bar n\cdot p n^\mu/2 + p_\perp^\mu$ is the label
momentum of the collinear quark $\xi_n$, which has momentum 
$p^\mu = \tilde p^{\mu} + k^{\mu}$, and we will set $p_\perp^\mu = 0$.
We are interested in the correlation function under the same
kinematic condition as the forward scattering amplitude of inclusive 
$B$ decays in the endpoint region. We denote the residual 
momentum $r$ as the remainder of the sum of large momenta of the
heavy quark, the collinear quark, and the incoming momentum $q$, with
\begin{equation} 
r^{\mu} = m_b v + q^{\mu} - \tilde{p}^{\mu} \sim \mathcal{O}(\Lambda),   
\quad n\cdot r = n\cdot q + m_b, \ \ \bar n\cdot r = r_\perp = 0.  
\end{equation}  
$n\cdot q$ is chosen to be negative, and $\Lambda$ is a typical
hadronic scale. Since $n\cdot r$ is of order $\Lambda$, the 
quark field in the intermediate state becomes hard-collinear with $p^2
\sim \mathcal{O} (m_b \Lambda)$. If we take a cut of the correlation
function $\Pi(q)$, the final state contains a pion or other collinear
particles which include a hard-collinear quark and an ultrasoft (usoft)
quark. The Feynman diagram for the correlation function at leading
order is schematically shown in Fig.~\ref{fig1}. 

The dispersion relation of the correlation function can be written as 
\begin{equation} 
\Pi [(p_B+q')^2] = \frac{1}{\pi} \int^{\infty}_{m_{\pi}^2 \to 0} 
d(p_B+q)^2 
\frac{\mathrm{Im} \Pi[(p_B+q)^2]}{(p_B+q)^2 - (p_B+q')^2 - i\epsilon}, 
\label{cor1} 
\end{equation}
where $p_B$ is the momentum of the $B$ meson. With $\oL = m_B -m_b$,
we can write 
\begin{eqnarray}
(p_B+q)^2 &=& (m_B v + q)^2 = [(m_b + \oL) v + q]^2\nonumber\\
&=& \n\cdot p (n\cdot r +  \oL) + \mathcal{O}(\Lambda^2),
\label{Bcon}
\end{eqnarray}
and similarly $(p_B+q')^2 = \n\cdot p(n\cdot r' +\oL) +
\mathcal{O}(\Lambda^2)$. The pion mass is regarded as
$m_{\pi}^2 \sim \Lambda^2$. In terms of the residual momentum $n\cdot
r$, the dispersion relation is
\begin{equation} 
\Pi (n\cdot r') = \frac{1}{\pi} \int^{\infty}_{m_{\pi}^2/\bar{n}\cdot
  p-\oL} dn\cdot r 
\frac{\mathrm{Im} \Pi (n\cdot r) }{n\cdot r - n\cdot r' - i\epsilon}. 
\label{cor2}
\end{equation} 
In the lower bound of the integral, $m_{\pi}^2/\bar{n}\cdot p$ can be
neglected compared to $\oL$. 

Using the optical theorem, the correlation function in the timelike
region can be written as  
\begin{equation} 
2\, \mathrm{Im} \Pi (r) = \sum_X \int \mathcal{D}\prod_X e^{i q\cdot z} 
\langle \overline{B} | J_0^{\dagger}(z) | X \rangle 
\langle X | J_0(0) | \overline{B} \rangle 
= 2\, \mathrm{Im} \Pi^{S} + 2\, \mathrm{Im} \Pi^{C},
\end{equation}  
where the superscripts `$S$' and `$C$' mean the contributions from the
single-pion state and from the continuum state respectively. 
The imaginary part of the correlation function from the single-pion state
is given by
\begin{equation} 
2\, \mathrm{Im} \Pi^S = \int \frac{d^3 p_{\pi}}{(2\pi)^3}
\frac{1}{2E_{\pi}} (2\pi)^4 \delta( p_B+q -p_{\pi} )
\langle \overline{B} | J_0^{\dagger} | \pi_n \rangle 
\langle \pi_n | J_0 | \overline{B} \rangle,  
\end{equation}
where the matrix element between the $B$ meson and the energetic pion
$\pi_n$ is expressed in term of the heavy-to-light form factor 
\begin{equation} 
\langle \pi_n | J_0 | \overline{B} \rangle = \overline{n} \cdot
p~\zeta_P (\mu).  
\end{equation} 
Therefore the correlation function for the single pion in the timelike
region can be written as 
\begin{equation} 
2\, \mathrm{Im} \Pi^{S} = 2\pi ~\delta((p_B+q)^2 - m_{\pi}^2)
~(\overline{n} \cdot p)^2 \zeta_P^2 (\mu) \sim  2\pi ~\delta((p_B+q)^2)
~(\overline{n}\cdot p)^2  \zeta_P^2 (\mu).
\label{pis} 
\end{equation}  
Using Eqs.~(\ref{cor1}), (\ref{Bcon}), and (\ref{pis}), 
the dispersion relation Eq.~(\ref{cor2}) can be rewritten as  
\begin{equation} 
\Pi (n\cdot r') = \frac{\n\cdot p \,\zeta_P^2}{-\oL - n\cdot r' -
  i\epsilon} + \frac{1}{\pi} \int^{\infty}_{w_s^{\prime}} d w
\frac{\mathrm{Im} \Pi (w)}{w - n\cdot r'-i\epsilon},
\label{corr} 
\end{equation} 
where $w = n\cdot r$ and $w_s^{\prime} = 4m_{\pi}^2/\bar{n} \cdot p
-\oL$ is related to the onset of the continuum states, which we take
as $s_0 = 4m_{\pi}^2= \overline{n} \cdot p (w_s^{\prime}+\oL)$.  

At the parton level, the correlation function $\Pi (n\cdot r')$ can be
obtained from SCET. From the matching between $\rm{SCET_I}$ 
and $\rm{SCET_{II}}$, the time-ordered product of the heavy-to-light
current and its hermitian conjugate can be expressed in terms of the
jet function and the $B$ meson shape function in
$\rm{SCET_{II}}$. Starting from Eq.~(\ref{cor}), the correlation
function is
\begin{eqnarray} 
\Pi(n\cdot r') &=& - m_B \int \frac{d \n\cdot z}{4\pi} \int dn\cdot k  
e^{i(n\cdot r' - n\cdot k) \n\cdot z /2} J_P(n\cdot k)
\nonumber \\ 
\label{cor3} 
&& \times \langle \overline{B}_v | \overline{h} Y \Bigl( \frac{\n\cdot
  z}{2} \Bigr) \frac{\fms{n}}{2} Y^{\dagger} h\Bigl(0\Bigr) |
  \overline{B}_v \rangle, 
\end{eqnarray} 
where the jet function $J_P$ with $P=\n \cdot p$ is defined as 
\begin{equation} 
\langle 0 | \mathrm{T} W^{\dagger}\xi_n (z)~ \overline{\xi}W (0) |
0\rangle = i \frac{\fms{n}}{2} \delta(n\cdot z) \delta(z_{\perp})  
\int \frac{dn\cdot k}{2\pi} e^{-in\cdot k'\n\cdot z /2}  
J_P( n\cdot k).
\label{jet} 
\end{equation} 
The jet function can be computed perturbatively and at tree level it
is given by  
\begin{equation} 
J_P^{(0)} (n\cdot k) = \frac{1}{n\cdot k + i\epsilon}.
\end{equation}
The $B$ meson shape function is defined as 
\begin{eqnarray} 
\langle \overline{B}_v | \overline{h} Y \Bigl( \frac{\n\cdot z}{2} 
\Bigr)  Y^{\dagger} h\Bigl(0\Bigr) | \overline{B}_v \rangle &=& 
\int d n\cdot l e^{in\cdot l \n \cdot z/2} 
\langle \overline{B}_v | \overline{h} Y \delta(n\cdot l - n\cdot
i\partial) \Bigr)  Y^{\dagger} h | \overline{B}_v \rangle  
\nonumber \\ 
&=&\int d n\cdot l e^{in\cdot l \n \cdot z/2} f(n\cdot l) \mathrm{Tr} 
\frac{1+\fms{v}}{2} 
\label{shape} \nonumber  \\
&=&2 \int d n\cdot l e^{in\cdot l \n \cdot z/2} f(n\cdot l),  
\end{eqnarray} 
where the residual momentum $n\cdot l$ should be less than 
$\oL$ since the shape function has support for $n\cdot l \le \oL$. 
Because $n\cdot k = n\cdot l + n\cdot r'$ from the phase space integral 
in Eq.~(\ref{cor3}), the correlation function can be written as 
\begin{equation}
\Pi(n\cdot r') = - m_B \int^{\oL}_{-\infty} dn\cdot l 
f(n\cdot l) J_P(n\cdot l + n\cdot r' + i\epsilon). 
\end{equation} 
At tree level, the correlation function becomes
\begin{equation} 
\Pi (n\cdot r') = m_B \int^{\infty}_{-\oL} dw 
\frac{f(-w= n\cdot l)}{w-n\cdot r'-i\epsilon},
\label{cor4}
\end{equation}
where we have changed the integral variable to $w = - n\cdot l$ to
allow for an easier comparison with Eq.~(\ref{corr}).

From Eqs.~(\ref{corr}) and (\ref{cor4}), we obtain the tree-level
relation 
\begin{equation} 
m_B \int^{\infty}_{-\oL} dw \frac{f(-w)}{w-n\cdot r'-i\epsilon} 
= \frac{\n\cdot p\, \zeta_P^2}{-\oL - n\cdot r' - i\epsilon} 
+ \frac{1}{\pi} \int^{\infty}_{w_s^{\prime}} d w
\frac{\mathrm{Im} \Pi (w)}{w - n\cdot r'-i\epsilon}\,.
\label{cor5} 
\end{equation}
Taking a Borel transformation on each side of Eq.~(\ref{cor5}), we
obtain 
\begin{equation}
m_B \int_{-\oL}^{\infty} dw f(-w) e^{-w/w_M} = \overline{n} \cdot p
\zeta_P^2 e^{\oL/w_M} +\frac{1}{\pi} \int_{w_s^{\prime}}^{\infty} dw
\mathrm{Im} \, \Pi (w) e^{-w/w_M}.   
\end{equation}
From this relation, we can relate the heavy-to-light form factor to
the $B$ meson shape function as 
\begin{eqnarray}
\zeta_P^2 &=& \frac{1}{\overline{n} \cdot p} \Bigl[ m_B
\int_{-\oL}^{\infty} dw f(-w) e^{-(w+\oL)/w_M} -\frac{1}{\pi}
  \int_{w_s^{\prime}}^{\infty} dw \mathrm{Im}\, \Pi (w)
  e^{-(w+\oL)/w_M} \Bigr] \nonumber \\
&=& \frac{m_B}{\overline{n} \cdot p} \Bigl[ \int_{-\oL}^{\infty} dw
f(-w) e^{-(w+\oL)/w_M} -\int_{w_s^{\prime}}^{\infty} dw f(-w)
e^{-(w+\oL)/w_M} \Bigr] \nonumber \\
&=&  \frac{m_B}{\overline{n} \cdot p} \int_{-\oL}^{w_s^{\prime}} dw
f(-w) e^{-(w+\oL)/w_M}.
\end{eqnarray}
By changing the variable $w+\oL \rightarrow w$, we finally obtain our
main result 
\begin{equation} 
\zeta_P^2  = \frac{m_B}{\n\cdot p} \int^{w_s}_0
dw f(\oL-w)~ e^{-w/w_M},
\label{relation}
\end{equation} 
where the Borel parameter $w_M$ is given by $M^2 = w_M
\overline{n}\cdot p \sim m_b \Lambda$ with $w_M \sim \Lambda_{\rm
  QCD}$. The other parameter $w_s$ is given by $w_s = w^{\prime}_s + \oL$
with $s_0 = 4m_{\pi}^2 = w_s \overline{n} \cdot p$.  Note that $w_s$
is a measure of the distance to the continuum and is roughly of
order $\Lambda_{\rm QCD}^2/m_b$.

When we expand the integrand in a Taylor
series near $w=0$, Eq.~(\ref{relation}) becomes 
\begin{eqnarray}\label{zeta2}
\zeta_P^2 &=& \frac{m_B}{\n\cdot p} \int^{w_s}_0 dw
\left\{f(\oL) - w\left[f'(\oL) + \frac{f(\oL)}{w_M}\right] +
  \frac{w^2}2\left[f''(\oL) + \frac{2 f'(\oL)}{w_M} +
    \frac{f(\oL)}{w_M^2}\right]+\cdots\right\}\nonumber\\ 
\label{taylor}
&=& \frac{m_B}{\n\cdot p} 
\left\{ f(\oL) w_s - \frac{{w_s}^2}2\left[ f'(\oL) +
    \frac{f(\oL)}{w_M}\right] + \frac{{w_s}^3}6\left[f''(\oL) +
    \frac{2 f'(\oL)}{w_M} +
    \frac{f(\oL)}{w_M^2}\right]+\cdots\right\}. 
\end{eqnarray}
Since $f(\oL)=0$, Eq.~(\ref{zeta2}) simplifies to
\begin{equation} \label{zeta3}
 \zeta_P^2 = \frac{m_B}{\n\cdot p} \left\{  - \frac{{w_s}^2}2
 f'(\oL)  + \frac{{w_s}^3}6\left[f''(\oL) +
    \frac{2 f'(\oL)}{w_M} \right]+\cdots\right\}. 
\end{equation}
>From various arguments \cite{chernyak,Beneke:2000ry} the soft form
factor $\zeta_P$ scales as $(\Lambda_{\rm QCD}/m_b)^{3/2}$, which puts
a constraint on the behavior of the $B$ meson shape function
near $\oL$, though the functional form is unknown. 
Since the left-hand side scales as $(\Lambda_{\rm QCD}/m_b)^3$, we can
constrain the scaling of the terms on the right-hand side.  
With $w_s \sim \Lambda^2/m_b$, the shape function $f(\oL)$ and its
derivatives of $f(\oL)$ should scale as $f(\oL) \sim \Lambda$,
$f'(\oL) \sim 1/\Lambda$, $f''(\oL) \sim 1/\Lambda^3$,  $\cdots$,
based on Eq.~(\ref{zeta2}). In general the shape function has
a width of order $\oL$ and is normalized to one, so it has a typical 
size of $1/\oL \sim 1/\Lambda$. There is an additional
factor of $1/\Lambda$ for each derivative of the shape function, from
which we expect the naive scaling like $f(\oL) \sim 1/\Lambda$,
$f^{\prime}(\oL) \sim 1/\Lambda^2$. The constraint from the scaling
behavior of the soft form factor does not allow this naive scaling. In
fact, $f(\oL)$ is suppressed by $\Lambda^2$ and $f^{\prime}(\oL)$ is
suppressed by $\Lambda$ compared to the naive scaling 
behavior due to this constraint. Practically we put $f(\oL)=0$ since
it is suppressed by $\Lambda^2$ compared to the naive scaling
behavior, and assume that the series in 
Eq.~(\ref{zeta3}) converges rapidly such that we choose the terms with
the first derivative. However this scaling behavior, especially
$f^{\prime} (\oL) \sim 1/\Lambda$ puts a constraint on the endpoint
behavior for the shape function models. 

\begin{figure}[b]
\begin{center}
\epsfig{file=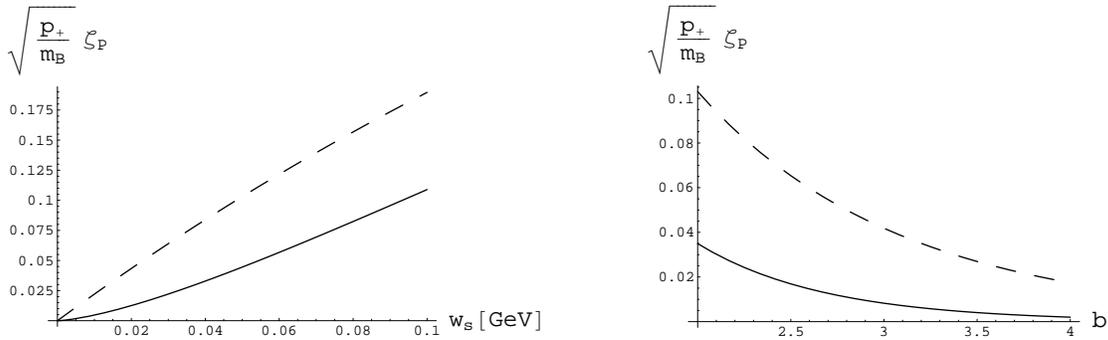, width=16cm}
\end{center}
\vspace{-1.0cm}
\caption{Dependence of $\zeta_P$ on the input parameter
  $w_s$ and $b$. In the first plot, the solid line denotes the soft
  form factor with $b$ = 2.93 and the dashed line with $b$ =
  2.00. In the second plot, the solid line represents the
  variation on $b$ with $w_s = 0.016$ GeV and the dashed line with
  $w_s = 0.05$ GeV.} 
\label{fig2}
\end{figure}

We can estimate the size of $\zeta_P$ from Eq.~(\ref{relation}) given
a model of the shape function.  We adopt the model of
Ref.~\cite{shapemodel}, which is given as 
\begin{equation}
f(\omega) = \frac{b^b}{\oL \Gamma(b)}
\left(1-\frac{\omega}{\oL}\right)^{b-1} e^{b (\omega/\oL-1)},
\label{model}
\end{equation}
at leading order in $\alpha_s$.   
The constraint that the first derivative scales as $1/\Lambda_{\rm
  QCD}$ implies that $b > 2$ for this model.  The default choice of
parameters in Ref.~\cite{shapemodel} is $\oL = 0.63$ GeV and $b =
2.93$, consistent with this constraint.  With this choice of
the parameters, and using $w_s = 4 m_\pi^2/m_b = 0.016$ GeV, $w_M =
0.5$ GeV, we obtain 
\begin{equation}
\zeta_P \approx 0.01 \sqrt{\frac{m_B}{\overline{n}\cdot p}}. 
\end{equation}
On the other hand, if we choose $w_s$ as 0.05 GeV, which can be
considered as a typical scale $\Lambda_{\rm QCD}^2/m_b$ with
$\Lambda_{\rm QCD} \sim 0.5$ GeV, we get
\begin{equation}
\zeta_P \approx 0.045 \sqrt{\frac{m_B}{\overline{n}\cdot p}}.  
\end{equation}

The behavior of $\zeta_P$ with respect to $w_s$ and $b$ is illustrated
in Fig.~\ref{fig2}. The soft form factor $\zeta_P$ with the input
parameter $b=2.93$ is smaller than 0.1 in the region $w_s = [0, 0.1]$
GeV and the central value is $\zeta_P \sim 0.05$
However, when we set the input parameter $b = 2.0$, the soft form
factor can be enhanced almost three times compared with the case of $b
= 2.93$. This can be easily understood when we consider
Eqs.~(\ref{zeta3}) and (\ref{model}). In the case with $b=2$, the
shape function in Eq.~(\ref{model}) scales as $f^{\prime}(\oL)\sim
1/\Lambda^2$, which makes $\zeta_P \sim \Lambda$ overestimating the
scaling behavior. The shape function model in 
Eq.(\ref{model}) with $b\sim 3$ is reliable considering the
scaling behavior of $\zeta_P$ near the tail of the shape function.

\begin{figure}[b]
\begin{center}
\epsfig{file=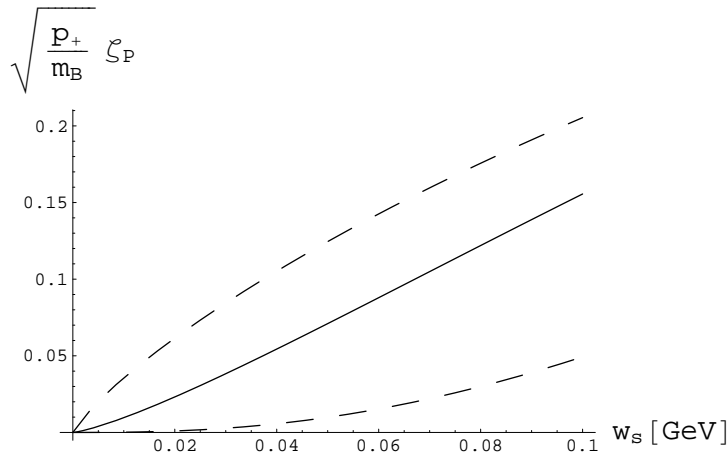, width=10cm}
\end{center}
\vspace{-1.0cm}
\caption{Tree-level estimate of $\zeta_P$ with the fitted input parameters 
extracted from the $B\to X_s \gamma$ data. The solid line shows the 
soft form factor with the central fitted parameters
($\overline{\Lambda}=0.545$ GeV, $\lambda_1 = -0.342
~\mathrm{GeV^2}$).  
The two dashed lines represent the range of the soft form factor 
with $1\sigma$ error, where the upper line is obtained with the input
parameters ($\overline{\Lambda}=0.781$ GeV, $\lambda_1 = -1.13
~\mathrm{GeV^2}$), and the lower line with $\overline{\Lambda}=0.485$ 
GeV, $\lambda_1 = -0.13 ~\mathrm{GeV^2}$.}     
\label{fig3}
\end{figure}
We can estimate the size of the soft form factor using the shape function
model extracted from $B\to X_s\gamma$ \cite{Bornheim:2002du}.
The central
fitted values on the exponential model of the shape function in
Eq.~(\ref{model}) are  
\begin{equation} 
\overline{\Lambda} = 0.545 ~\mathrm{Gev},~~~~~
\lambda_1 = -0.342 ~\mathrm{GeV^2},
\label{cent} 
\end{equation} 
where $\lambda_1$ is defined as $\langle \overline{B}_v | \overline{h}_v
(n\cdot iD_{us})^2 h_v |\overline{B}_v \rangle = - \lambda_1/3$ and,
using the relation \cite{shape}   
\begin{equation} 
\int d\omega ~\omega^2 f(\omega) =\frac{\overline{\Lambda}^2}{b} =
-\frac{\lambda_1}{3},  
\end{equation}    
we can obtain the value of the input parameter $b=2.605$ under the choice 
of Eq.~(\ref{cent}). 
Using the fitted shape function from the data on $B \to X_s \gamma$, 
the soft form factor at tree level with $1\sigma$ error is given by 
\begin{eqnarray} 
\sqrt{\frac{\bar{n} \cdot p}{m_B}} \zeta_P =\! \bigg\{ \begin{array}{l}
 0.0175^{+0.0342}_{-0.0167} ~~(w_s = 0.016 ~\mathrm{GeV}), \\ 
 0.0710^{+0.0540}_{-0.0580}~~ (w_s = 0.05  ~\mathrm{GeV}). \end{array} 
\end{eqnarray} 
The behavior of $\zeta_P$ using this shape function model is shown in
Fig.~\ref{fig3}. In the heavy-to-light soft form factor such as
$\zeta_P$, it is reasonable to choose $w_s$ as $4m_{\pi}^2/m_b$ since
the invariant mass squared of the lowest continuum state starts at $p^2
\sim 4m_{\pi}^2$. Therefore, with this choice of $w_s$, we find 
$\zeta_P(B\to\pi)$ at tree level can be small contrary to the prediction of 
the QCD factorization approach \cite{QCDF,QCDFhl}.

The numerical analysis of the soft form factor $\zeta_P$ is based on
the tree-level relation in Eq.~(\ref{relation}). There can be a few
sources of theoretical uncertainties. For example, there may be
uncertainties in choosing the Borel parameters $w_s$ and
$w_M$. Since the forward scattering is computed in
$\mathrm{SCET}_{\mathrm{I}}$, the virtuality of the intermediate
state, or the pion state is of order $M^2 =w_M \overline{n}\cdot p\sim
m_b \Lambda$, and $w_M \sim \Lambda$. And $w_s$ is the scale from
which the continuum states start. We take as $w_s =
4m_{\pi}^2/\overline{n}\cdot p$ or $\sim \Lambda^2/m_b$ with $\Lambda
\approx 0.5$ GeV, assuming that
$m_{\pi}^2 \sim \Lambda^2$, which is obviously an overestimation. This
is in contrast to the choice of the Borel parameters in
Ref.~\cite{DeFazio:2005dx}, in which they take $w_s \approx w_M
\approx 0.2$ GeV. Their tree-level evaluation of the soft form factor
corresponds to $\zeta_P \sim 0.32$, while our evaluation is about 0.01
($w_s = 0.016$ GeV) and 0.05 ($w_s =0.05$ GeV), which is smaller by an
order of magnitude. We find that the variation of $w_M$ between 0.2
GeV and 0.6 GeV gives 1\% ($w_s=0.016$ GeV), 10\% ($w_s=0.05$
GeV). Therefore we can argue that the large uncertainty arises from
the choice of $w_s$. Indeed, we obtain $\zeta_P \approx 0.30$ in our
evaluation with $w_s =0.2$ GeV, $w_M=0.4$ GeV, close to the result in
Ref.~\cite{DeFazio:2005dx}.    

Let us briefly comment on the higher-order corrections to the
tree-level relation in Eq.~(\ref{relation}). First, we have to
include the radiative correction to $\Pi (q)$, which include the
radiative corrections of $J_0$. In addition, we also have to include
the spectator interactions. 
When we consider a hard collinear gluon exchange with the spectator quark, 
the contributions can be separated into two parts, 
the soft form factor $\zeta_P$ and the hard form factor $\zeta_J$,   
based on the existence of the spin symmetry of the heavy-to-light
current \cite{Beneke:2000ry,QCDFhl}. The hard spectator contributions
for $\zeta_P$ satisfying the spin symmetry can be 
distinguished simply by the presence of the leading order current
$J_0$ in the time-ordered product \cite{Bauer:2002aj}. 
In this case we need to include the higher order 
interaction terms of $\rm{SCET_I}$ collinear 
Lagrangian in the correlation function. 

\begin{figure}[t]
\begin{center}
\epsfig{file=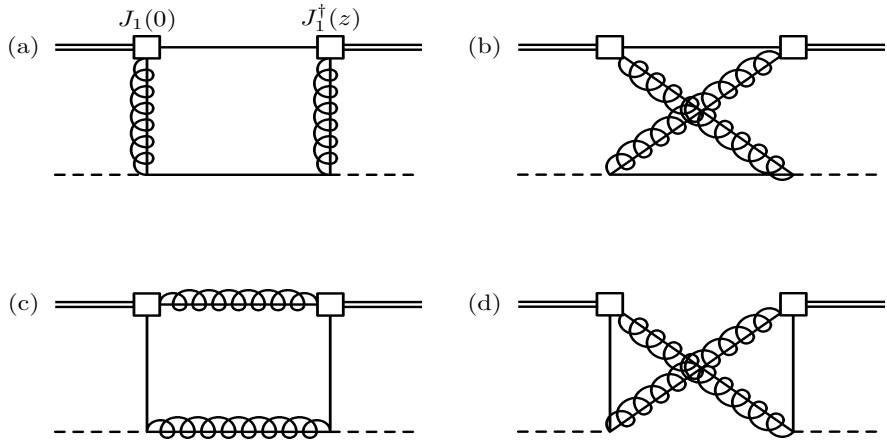, width=12cm, height=7cm}
\end{center}
\vspace{-1.0cm}
\caption{Feynman diagrams for the lowest-order contributions to
  $\Pi_1(q)$ in $\rm{SCET_I}$.} 
\label{fig4}
\end{figure}

\begin{figure}[t]
\begin{center}
\epsfig{file=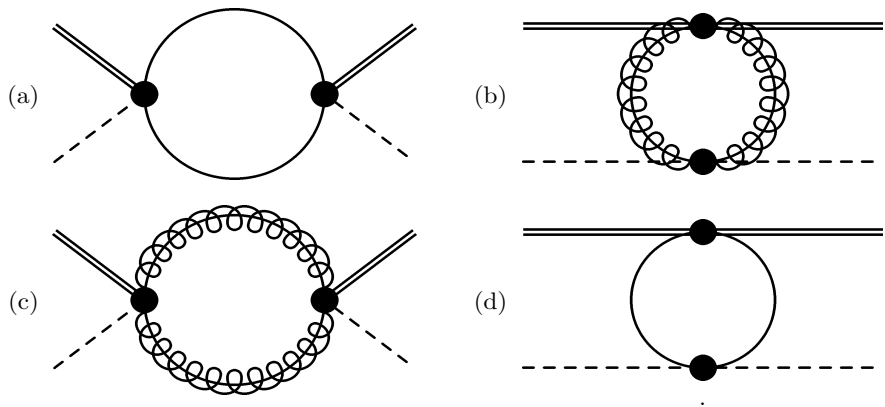, width=12cm, height=6cm}
\end{center}
\vspace{-1.0cm}
\caption{Examples on the nonlocal time-ordered products in $\rm{SCET_{II}}$.
The internal loops consist of the collinear particles with offshellness 
$\Lambda^2$. The dotted vertices denote the four-particle operators  
obtained from the matching respectively 
by integrating out the hard-collinear gluons (a,d) and quark fields
(b,c).} 
\label{fig5}
\end{figure}

The spin symmetry breaks down explicitly if subleading currents with 
$i\fmsl{D}_c^{\perp}$ are included in the time-ordered product. 
Therefore,  
when the hard spectator contributions for $\zeta_J$ are considered,
we have to define a
correlation function with the subleading current $J_1$, which is given
by 
\begin{equation}
 J_1 (x) = e^{i(\tilde{p}-m_b v)\cdot x} \overline{\xi}_n
 \frac{\fms{\overline{n}}}{2} i{\fmsl{D}^{\perp}_c}
WY^{\dagger} h_v (x).  
\end{equation}
In this case, the intermediate state consists of hard-collinear
particles and one of the correlation functions is given by
\begin{equation}
\Pi_1 (q^2) = i\int d^4 z e^{iq\cdot z} \int d^4x \int d^4 y \langle
\overline{B} | T\Bigl[J_1^{\dagger} (z) J_1 (0) i\mathcal{L}_{\xi
  q}^{(1)} (x) i\mathcal{L}_{\xi q}^{(1)} (y) \Bigr]
|\overline{B}\rangle, 
\end{equation}
where $\mathcal{L}_{\xi q}^{(1)} = \overline{\xi}_n
  (\fms{\mathcal{P}}^{\perp} +g\fms{A}_n^{\perp}) WY^{\dagger} q_{us}$
  is the ultrasoft-collinear 
Lagrangian. This correlation function contributes to the hard form
factor. After integrating out the hard-collinear degrees of
freedom, we have four-quark operators with $h_v$ and $q_{us}$. 
As seen in Fig.~\ref{fig4}, 
from
the four possible contractions of the hard-collinear particles in the
intermediate state 
we obtain the operators in
$\mathrm{SCET}_{\mathrm{II}}$ of the form 
\begin{eqnarray}
 O_1 (l_1^+, l_2^+) &=& \bigl(\overline{h} Y\bigr)_a
 \frac{\fms{\overline{n}}}{2} \delta (l_1^+-n\cdot i\partial)
 \bigl(Y^{\dagger} h \bigr)_b
 \cdot \bigl(\overline{q}_{us} Y\bigr)_b \frac{\fms{n}}{2}
 \delta (l_2^+-n\cdot i\partial)\bigl(Y^{\dagger} q_{us} \bigr)_a,
 \\ 
 O_2 (l_1^+, l_2^+) &=&  \bigl(\overline{h} Y\bigr)_a
 \frac{\fms{\overline{n}}}{2} 
 \sigma_{\perp    \alpha \beta}  \delta (l_1^+-n\cdot i\partial)\bigl(
 Y^{\dagger} h\bigr)_b  \cdot \bigl(\overline{q}_{us} Y\bigr)_b
 \frac{\fms{n}}{2}  \sigma_{\perp}^{\alpha\beta}  \delta (l_2^+-n\cdot
 i\partial)\bigl(Y^{\dagger} q_{us}\bigr)_a, 
 \nonumber \\ 
O_3 (l_1^+, l_2^+)  &=& \bigl(\overline{h} Y\bigr)_a
\frac{\fms{\overline{n}}   \fms{n}}{4}  \delta (l_1^+-n\cdot
i\partial)\bigl(Y q_{us} \bigr)_a \cdot \bigl( \overline{q}_{us} Y)_b
\frac{\fms{n} \fms{\overline{n}}}{4}  \delta (l_2^+-n\cdot
i\partial)\bigl(Y^{\dagger} h\bigr)_b, \nonumber \\
O_4  (l_1^+, l_2^+)&=& \bigl(\overline{h} Y\bigr)_a
\frac{\fms{\overline{n}}   \fms{n}}{4} \sigma_{\perp    \alpha \beta}
\delta (l_1^+-n\cdot i\partial)\bigl(Y q_{us} \bigr)_a \cdot \bigl(
\overline{q}_{us}  Y)_b \frac{\fms{n} \fms{\overline{n}}}{4}
\sigma_{\perp}^{\alpha\beta} \delta (l_2^+-n\cdot
i\partial)\bigl(Y^{\dagger} h\bigr)_b, \nonumber 
\end{eqnarray}
where $\sigma_{\perp}^{\alpha\beta} = i[\gamma_{\perp}^{\alpha},
\gamma_{\perp}^{\beta}]/2$, and $a$, $b$ are color indices.  
In addition, we also have to consider the possibilities 
that the hard-collinear 
loop diagrams in Fig.~\ref{fig4} can be matched onto the nonlocal time-ordered
products such as Fig.~\ref{fig5} in $\rm{SCET_{II}}$. When calculating 
the hard-collinear loop corrections, we are confronted with non-vanishing
infrared divergences which result from internal propagators with virtualities  
of order $\Lambda^2$. In this case we expect 
the infrared divergences will be reproduced by the possible nonlocal 
time-ordered products in $\rm{SCET_{II}}$.

Therefore, by matching onto $\mathrm{SCET}_{\mathrm{II}}$, we obtain
\begin{eqnarray} 
\Pi_1 &=& \int dl_1^+ dl_2^+ 
\sum_{i=1}^4 J_i (\overline{n}\cdot p,
n\cdot r, l_1^+, l_2^+) \langle \overline{B}|O_i (l_1^+, l_2^+)
|\overline{B} \rangle \nonumber \\
&&+ \int d^4 x d^4 y
\sum_{i,k}  J'_i J'_k \otimes \langle \overline{B}| \mathrm{T} 
[O_i^L(x), O_k^L(y)]|\overline{B} \rangle + \cdots,  
\label{hard1}
\end{eqnarray}
where $J_i,J'_i$ are the jet functions and 
$\otimes$ denotes convolutions of $l_i^+,l_k^+$ which 
can appear in the operators $O_i^L, O_k^L$. 
The matrix elements between the $B$
meson in Eq.~(\ref{hard1}) are new nonperturbative parameters to be
included by defining additional subleading shape
functions of the $B$ meson. 
Similarly, 
there are also hard spectator contributions to $\zeta_P$
from other time-ordered products with the subleading
collinear Lagrangian, which introduce more nonperturbative parameters.   
However, we have no information on these subleading $B$ meson
shape functions, and hence we do not further consider
higher-order corrections and other theoretical uncertainties such as
the renormalization scale dependence in this paper. 

We have derived the tree-level relation between the
soft form factor for $B\rightarrow \pi$ and the leading $B$ meson shape function
using the sum rule approach. This is achieved by
considering the forward scattering amplitude of the heavy-to-light
current near the endpoint. We derived a constraint on the behavior of the
$B$ meson shape function near the endpoint using the scaling
behavior of the soft form factor, $\zeta_P \sim
(\Lambda/m_b)^{3/2}$. Since we work at tree level only, there may be a
lot of uncertainties, and there may also be a model dependence on the
form of the shape function. But the analysis suggests that the soft form 
factor can be about 0.01 to 0.07, considering various inputs from the
models and using the experimental data on the $B$ meson shape function. 
The numerical value of the soft form
factor is approximately smaller by an order of magnitude than other sum rule
predictions. Since we have not included higher-order corrections in
$\alpha_s$, this evaluation may have a large theoretical
uncertainty, but our analysis favors a small value of the soft form
factor. 

\section*{Acknowledgments}
We thank Ed Thorndike for his help in providing the parameters  
for the experimentally extracted shape function.
We also thank the Institute for Nuclear Theory at the University of
Washington for its hospitality and the Department of Energy for
partial support during the completion of this work. 
J.~Chay was supported by Grant No. R01-2002-000-00291-0 from the Basic
Research Program of the Korea Science \& Engineering
Foundation. C.~Kim and A~.K.~Leibovich were supported by 
the National Science Foundation under Grant No. PHY-0244599.
A.~K.~Leibovich was also supported in part by the Ralph E.~Powe Junior
Faculty Enhancement Award.


\begin{thebibliography}{99}
\normalsize
\baselineskip 3.0ex 

\bibitem{gammaextraction}
  C.~W.~Bauer, I.~Z.~Rothstein and I.~W.~Stewart,
  Phys.\ Rev.\ Lett.\  {\bf 94} (2005) 231802;
  M.~Beneke, G.~Buchalla, M.~Neubert and C.~T.~Sachrajda,
  arXiv:hep-ph/0411171;
  C.~W.~Bauer, D.~Pirjol, I.~Z.~Rothstein and I.~W.~Stewart,
  arXiv:hep-ph/0502094.

\bibitem{lattice}
  M.~Okamoto {\it et al.},
  Nucl.\ Phys.\ Proc.\ Suppl.\  {\bf 140} (2005) 461;
  J.~Shigemitsu {\it et al.},
  arXiv:hep-lat/0408019.

\bibitem{lcsr}
  P.~Ball and R.~Zwicky,
  JHEP {\bf 0110} (2001) 019;
  Phys.\ Rev.\ D {\bf 71} (2005) 014015.

\bibitem{DeFazio:2005dx}
  F.~De Fazio, T.~Feldmann and T.~Hurth,
  arXiv:hep-ph/0504088.

\bibitem{Ball:2003bf}
  P.~Ball,
  arXiv:hep-ph/0308249.
  
\bibitem{scet}
C.~W.~Bauer, S.~Fleming and M.~E.~Luke,
Phys.\ Rev.\ D {\bf 63} (2001) 014006;
C.~W.~Bauer, S.~Fleming, D.~Pirjol and I.~W.~Stewart,
Phys.\ Rev.\ D {\bf 63} (2001) 114020;
C.~W.~Bauer, D.~Pirjol and I.~W.~Stewart,
Phys.\ Rev.\ D {\bf 65} (2002) 054022.

\bibitem{shape}
  M.~Neubert,
  Phys.\ Rev.\ D {\bf 49} (1994) 3392;
  T.~Mannel and M.~Neubert,
  Phys.\ Rev.\ D {\bf 50} (1994) 2037.
  
\bibitem{Bornheim:2002du}
  A.~Bornheim {\it et al.}  [CLEO Collaboration],
  Phys.\ Rev.\ Lett.\  {\bf 88} (2002) 231803.
  
\bibitem{Bauer:2002aj}
  C.~W.~Bauer, D.~Pirjol and I.~W.~Stewart,
  Phys.\ Rev.\ D {\bf 67} (2003) 071502.

\bibitem{chernyak} V.~L.~Chernyak and I.~R.~Zhitnitsky, Nucl.\ Phys.\
  B {\bf 345} (1990) 137.

\bibitem{Beneke:2000ry}
  M.~Beneke, G.~Buchalla, M.~Neubert and C.~T.~Sachrajda,
  Nucl.\ Phys.\ B {\bf 591} (2000) 313.


\bibitem{shapemodel}
  S.~W.~Bosch, B.~O.~Lange, M.~Neubert and G.~Paz,
  Nucl.\ Phys.\ B {\bf 699} (2004) 335;
S.~W.~Bosch, M.~Neubert and G.~Paz,
  JHEP {\bf 0411} (2004) 073.

\bibitem{QCDF}
  M.~Beneke, G.~Buchalla, M.~Neubert and C.~T.~Sachrajda,
  Phys.\ Rev.\ Lett.\  {\bf 83} (1999) 1914.

\bibitem{QCDFhl}  
  M.~Beneke and T.~Feldmann,
  Nucl.\ Phys.\ B {\bf 592} (2001) 3.


\end{thebibliography}
\end{document}